\documentclass{ws-rv9x6}

\usepackage{ws-rv-van}             

\usepackage{amsmath}
\usepackage{amsfonts}
\usepackage{mathrsfs}

\makeindex

\newcommand{\beqs}{\begin{equation*}}
\newcommand{\beq}{\begin{equation}}

\newcommand{\eeqs}{\end{equation*}}
\newcommand{\eeq}{\end{equation}}

\newcommand{\beqas}{\begin{eqnarray*}}
\newcommand{\beqa}{\begin{eqnarray}}

\newcommand{\eeqas}{\end{eqnarray*}}
\newcommand{\eeqa}{\end{eqnarray}}

\newcommand{\eq}[2]{\begin{equation} #1 \label{#2} \end{equation}}

\newcommand{\eps}{\varepsilon}
\newcommand{\al}{\alpha}
\newcommand{\be}{\beta}

\newcommand{\de}{\delta}
\newcommand{\om}{\omega}

\newcommand{\la}{\lambda}
\newcommand{\si}{\sigma}

\newcommand{\Ga}{\Gamma}

\newcommand{\blist}{\begin{itemize}}

\newcommand{\elist}{\end{itemize}}

\providecommand{\href}[2]{#2}

\DeclareFontFamily{OT1}{rsfs}{}
\DeclareFontShape{OT1}{rsfs}{m}{n}{ <-7> rsfs5 <7-10> rsfs7 <10->rsfs10}{} 
\DeclareMathAlphabet{\mycal}{OT1}{rsfs}{m}{n}

\begin{document}


\chapter{Canonical analysis of cosmological topologically massive gravity at the chiral point\label{ch1}}

\author[D.~Grumiller, R.~Jackiw and N.~Johansson]{Daniel Grumiller\footnote{E-mail: {\tt grumil@hep.itp.tuwien.ac.at}}, Roman Jackiw\footnote{E-mail: {\tt jackiw@lns.mit.edu}} and Niklas Johansson\footnote{E-mail: {\tt Niklas.Johansson@fysast.uu.se}}}

\address{
\footnotemark[1]\footnotemark[2] Massachusetts Institute of Technology, \\
77 Massachusetts Ave., Cambridge, MA 02139 \\
\footnotemark[3] Institutionen f\"{o}r fysik och astronomi, Uppsala universitet, \\
Box 803, S-751 08 Uppsala, Sweden
}

\begin{abstract}
Wolfgang Kummer was a pioneer of two-dimensional gravity and a strong advocate of the first order formulation in terms of Cartan variables. In the present work we apply Wolfgang Kummer's philosophy, the `Vienna School approach', to a specific three-dimensional model of gravity, cosmological topologically massive gravity at the chiral point. Exploiting a new Chern--Simons representation we perform a canonical analysis. The dimension of the physical phase space is two per point, and thus the theory exhibits a local physical degree of freedom, the topologically massive graviton.
\end{abstract}

\body


\section{Introduction}

Gravity in lower dimensions provides an excellent expedient for testing ideas about classical and quantum gravity in higher dimensions. The lowest spacetime dimension where gravity can be described is two, and Wolfgang Kummer contributed significantly to research on two-dimensional gravity, see Ref.~\cite{Grumiller:2002nm} for a review. Those who knew Wolfgang will recall that one of his main points was to advocate a gauge theoretic approach towards gravity, see Ref.~\cite{Kummer:2005tx} for his last proceedings contributions. Instead of using the metric, $g_{\mu\nu}$, as fundamental field he insisted on employing the Cartan variables, Vielbein $e^a_\mu$ and connection $\om^a{}_{b\,\mu}$. His approach greatly facilitated the canonical analysis and the quantization of the theory.

In the present work we shall study gravity in three dimensions along similar lines. We start by collecting a few well-known features of gravity in three dimensions. Pure Einstein--Hilbert gravity exhibits no physical bulk degrees of freedom \cite{Weinberg:1972,Deser:1984tn,Deser:1984dr}. If the theory is deformed by a negative cosmological constant it has black hole solutions \cite{Banados:1992wn}. Another possible deformation is to add a gravitational Chern--Simons term. The resulting theory is called topologically massive gravity (TMG) and, remarkably, contains a massive graviton \cite{Deser:1982vy}. 
Including both terms yields cosmological topologically massive gravity \cite{Deser:1982sv} (CTMG), a theory that exhibits both gravitons and black holes. Parameterizing the negative cosmological constant by $\Lambda=-1/\ell^2$ the (second order) action is given by 
\eq{
I_{\rm CTMG}[g]= \int d^3x\sqrt{-g}\,\Big[R+\frac{2}{\ell^2} 
+\frac{1}{2\mu} \,\eps^{\la\mu\nu}\Ga^\rho{}_{\la\si}\,\big(\partial_\mu \Ga^\si{}_{\nu\rho}+\frac23 \,\Ga^\si{}_{\mu\tau}\Ga^\tau{}_{\nu\rho}\big)\Big] \,.
}{eq:2nd}
In Ref.~\cite{Li:2008dq} it was advocated to study the theory \eqref{eq:2nd} at the chiral point
\eq{
\mu\ell = 1\,,
}{eq:chiral}
where the theory exhibits very special properties. We abbreviate this theory by the acronym CCTMG (`chiral cosmological topologically massive gravity'). By imposing the Brown--Henneaux boundary conditions Ref.~\cite{Li:2008dq} argued that CCTMG exhibits no bulk degrees of freedom. On the other hand Ref.~\cite{Carlip:2008jk} found that CCTMG exhibits one bulk degree of freedom. By slightly relaxing the Brown--Henneaux boundary conditions --- still requiring spacetime to be asymptotically AdS --- Ref.~\cite{Grumiller:2008qz} demonstrated that indeed a physical degree of freedom exists in CCTMG: the topologically massive graviton. The analyses in Refs.~\cite{Li:2008dq,Carlip:2008jk,Grumiller:2008qz} were focused on the linearized level, i.e., perturbing around an AdS$_3$ background.

In the present work we go beyond the linearized approximation and perform a non-perturbative (classical) canonical analysis of CCTMG (see also Refs.~\cite{Deser:1991qk,Park:2008yy,Carlip:2008}).\footnote{For further recent literature related to CCTMG see Refs.~\cite{Banerjee:2008ez,Compere:2008us,Alishahiha:2008tv,Hotta:2008yq,Li:2008yz,Sachs:2008gt,Lowe:2008ye}.} Our main goal is to derive the dimension of the physical phase space, which allows us to deduce the number of physical bulk degrees of freedom.

This paper is organized as follows. In Section \ref{se:2} we present a new Chern--Simons formulation of cosmological topologically massive gravity. In Section \ref{se:3} we focus on the chiral point and establish the Hamiltonian formulation, identifying all primary, secondary and ternary constraints. In Section \ref{se:4} we perform a constraint analysis and check the first/second class properties of all constraints, which allows us to establish the dimension of the physical phase space. In Section \ref{se:5} we conclude.

Our conventions are as follows. We use Greek spacetime indices and Latin frame indices. The former are raised and lowered with the spacetime metric $g_{\mu\nu}$ and the latter with the flat metric $\eta_{ab}$. Both have signature $-,+,+$. For the Dreibein $e_\mu^a$ we choose ${\rm sign\,}(\det{e}) = 1$. When writing $p$-forms we usually suppress the spacetime indices, e.g.~$e^a$ denotes the 1-form $e^a=e^a_\mu dx^\mu$. We disregard boundary terms in the present work, so equivalences between actions have to be true only up to total derivatives.

\section{Chern--Simons formulation}\label{se:2}

Instead of the action \eqref{eq:2nd} which functionally depends on the metric one can equivalently use the action
\eq{
I_{\rm CTMG}[e]=\int \Big[2e^a \wedge R_a(\om) + \frac{1}{3\ell^2}\,\eps_{abc}\,e^a\wedge e^b\wedge e^c-\frac1\mu\, {\rm CS}(\om)\Big]
}{eq:can0}
which functionally depends on the Dreibein. The gravitational Chern--Simons term
\eq{
{\rm CS}(\om):=\om^a\wedge d\om_a + \frac13\,\eps_{abc}\,\om^a\wedge\om^b\wedge\om^c
}{eq:can2}
and the (dualized) curvature 2-form 
\eq{
R_a(\om):=d\om_a + \frac12\,\eps_{abc}\,\om^b\wedge \om^c\,
}{eq:can4}
depend both exclusively on the (dualized) connection defined by $\om^a:=\frac12\eps^{abc}\om_{bc}$. Note that the connection is not varied independently in the formulation \eqref{eq:can0}, but rather it is the Levi-Civita connection, i.e., metric compatible $\om^{ab}=-\om^{ba}$ and torsion-free, $T^a=0$, where
\eq{
T_a:=d e_a + \eps_{abc}\,\om^b\wedge e^c
}{eq:can3}
is the torsion 2-form. This means that $\om^a$ in \eqref{eq:can0} has to be expressed in terms of $e^a$ (and derivatives thereof) before variation.

For our purposes it is very convenient to employ a formulation where we can vary independently the Dreibein and the connection \cite{Baekler:1991}. This is achieved by supplementing the action \eqref{eq:can0} with a Lagrange multiplier term enforcing the torsion constraint,
\eq{
I_{\rm CTMG}[e,\om,\la]=\int \Big[2e^a \wedge R_a + \frac{1}{3\ell^2}\,\eps_{abc}\,e^a\wedge e^b\wedge e^c-\frac1\mu\, {\rm CS}(\om)+\la^a\wedge T_a\Big].
}{eq:can1a}
The first order action \eqref{eq:can1a} is classically equivalent \cite{Baekler:1991} to the second order action \eqref{eq:2nd}. This can be shown as follows. Varying \eqref{eq:can1a} with respect to $\lambda_a$ and $\om_a$ establishes the condition of vanishing torsion \eqref{eq:can3} and an algebraic relation for $\la_a$, 
\eq{
\frac12\,\eps_{abc}\,\la^a\wedge e^b = \frac 1\mu\, R_c-T_c = \frac 1\mu\, R_c\,,
}{eq:can83}
in terms of Dreibein, connection and derivatives thereof. Thus, both $\om_a$ and $\la_a$ can be expressed in terms of the Dreibein, and first and second derivatives thereof. Varying \eqref{eq:can1a} with respect to the Dreibein and plugging into that equation the relations for $\la_a$ and $\om_a$ in terms of $e_a$ yields a set of third order partial differential equations in $e_a$. Using the defining relation between Dreibein and metric, $g_{\mu\nu} = e^a_\mu e^b_\nu \,\eta_{ab}$, finally establishes 
\eq{
G_{\mu\nu} + \frac{1}{\mu}\, C_{\mu\nu} = 0\,,
}{eq:EOM}
where 
\eq{
G_{\mu\nu} = R_{\mu\nu} - \frac12\, g_{\mu\nu} R -\frac{1}{\ell^2}\, g_{\mu\nu}
}{eq:cg45}
is the Einstein tensor (including cosmological constant) and 
\eq{
C_{\mu\nu} = \frac12\,\eps_\mu{}^{\al\be} \,\nabla_\al R_{\be\nu} + (\mu\leftrightarrow\nu)
}{eq:cg58}
is essentially the Cotton tensor. The equations of motion \eqref{eq:EOM} also follow directly from varying the second order action \eqref{eq:2nd} with respect to the metric.

We make now some field redefinitions to further simplify the action \eqref{eq:can1a}. We shift the Lagrange multiplier $\la^a\to\la^a-e^a/(\mu\ell^2)$ and obtain
\eq{
I_{\rm CTMG}[e,\om,\la]=\int \Big[2e^a \wedge R_a + \frac{1}{3\ell^2}\,\eps_{abc}\,e^a\wedge e^b\wedge e^c-\frac1\mu\, {\rm CS}(\om)+\big(\la^a-\frac{e^a}{\mu\ell^2}\big)\wedge T_a\Big]
}{eq:can1}
In the absence of the $\la^a \wedge T_a$-term in \eqref{eq:can1}, the well-known field redefinitions
\eq{
A^a:=\om^a+e^a/\ell\,,\qquad \tilde A^a:=\om^a-e^a/\ell
}{eq:can5}
turn the action into a difference of two Chern--Simons terms \cite{Achucarro:1986vz,Witten:1988hc,Blagojevic:2003vn,Cacciatori:2005wz}. Curiously,
under the same redefinitions \eqref{eq:can5} the Lagrange multiplier term can be recast into a difference of two Einstein--Hilbert terms, where $\lambda$ plays the role of the Dreibein:
\eq{
\frac2\ell\, I_{\rm CTMG}[A,\tilde A,\la]= \big(1-\frac{1}{\mu\ell}\big)I_{\rm CS}[A]+I_{EH}[\la,A]-\big(1+\frac{1}{\mu\ell}\big)I_{\rm CS}[\tilde A]-I_{\rm EH}[\la,\tilde A].
}{eq:can6}
We have introduced here the abbreviations
\eq{
I_{\rm CS}[A]:=\int {\rm CS}(A)
}{eq:can7}
and
\eq{
I_{\rm EH}[\la,A]:=\int\la^a\wedge R_a(A)
}{eq:can8}
and similarly for $\tilde A$. 

The reformulation \eqref{eq:can6} of the action \eqref{eq:can1a} as difference of Chern--Simons and Einstein-Hilbert terms seems to be new. It is worthwhile repeating that in both Einstein--Hilbert terms the Lagrange multiplier $\la^a$ formally plays the role of a `Drei\-bein'. This suggests that $\la^a$ should be invertible. We have checked that for pure AdS$_3$ [which obviously solves the field equations \eqref{eq:EOM}] the symmetric tensor $\la_{\mu\nu} = e_{(\mu}^a \la_{\nu) \,a}$ is proportional to the metric. Thus, requiring invertibility of $\la^a$ is necessary in general to guarantee invertibility of the metric.

The advantage of the formulation \eqref{eq:can6} is twofold. Because the action contains only first derivatives (linearly) a canonical analysis is facilitated. Moreover, at the chiral point $\mu^2\ell^2=1$ one of the Chern--Simons terms vanishes. 

\section{Hamiltonian action at the chiral point}\label{se:3}

We focus now on the theory at the chiral point and assume for sake of specificity $\mu\ell=1$. The action \eqref{eq:can6} simplifies to
\eq{
I_{\rm CCTMG}[A,\tilde A,\la] = \frac{\ell}{2}\, I_{EH}(\la,A)-\ell\, I_{\rm CS}(\tilde A)-\frac{\ell}{2}\, I_{\rm EH}(\la,\tilde A)=\int d^3x\, {\cal L}
}{eq:can10}
To set up the canonical analysis one could now declare the 27 fields $\la^a$, $A^a$, $\tilde A^a$ to be canonical coordinates and calculate their 27 canonical momenta \cite{Park:2008yy}. In this way one produces many second class constraints which have to be eliminated by the Dirac procedure \cite{Dirac:1996}. However, this is not the most efficient way to start the canonical analysis. As realized in Ref.~\cite{Faddeev:1988qp} if an action is already in first order form a convenient short-cut exists. In the present case this short-cut consists basically of picking the appropriate sets of fields as canonical coordinates and momenta, respectively.

We use the 18 fields $\lambda_\mu^a,\tilde{A}_0^a,\tilde{A}_1^a,A_0^a$ as canonical coordinates and introduce the notation 
\eq{
q_1^a=\la_1^a\,,\;\, q_2^a=\la_2^a\,,\;\, q_3^a=\tilde A_1^a\,,\;\, \bar q_1^a=\la_0^a\,,\;\, \bar q_2^a = \tilde A_0^a\,,\;\, \bar q_3^a = A_0^a \,.
}{eq:notation}
Like in electrodynamics or non-abelian gauge theory the momenta $\bar p_i^a$ of the zero components $\bar q_i^a$ are primary constraints. The simplest way to deal with them is to exclude the pairs $\bar q_i^a,\bar p_i^a$ from the phase space and to treat the $\bar q_i^a$ as Lagrange multipliers for the secondary constraints (``Gauss constraints''). This reduces the dimension of our phase space to 18. The 9 momenta $p^a_i$, 
\begin{align}
\frac{\partial{\cal L}}{\partial\partial_0\la_{1\,a}} &= p^a_1 = \frac\ell 2 (A_2^a-\tilde A_2^a)=e_2^a\\
\frac{\partial{\cal L}}{\partial\partial_0\la_{2\,a}} &= p^a_2 = -\frac\ell 2 (A_1^a-\tilde A_1^a)=-e_1^a\\
\frac{\partial{\cal L}}{\partial\partial_0\tilde A_{1\,a}} &= p^a_3 = -2\ell\,\tilde A_2^a 
\end{align}
depend linearly on the fields $A_1^a,A_2^a,\tilde A_2^a$. These fields are not 
contained in our set of canonical coordinates.

The Hamiltonian action is now determined as
\eq{
I_{\rm CCTMG}[q,\,p;\,\bar q] = \int d^3x \,\big(p_{i\,a} \dot q^a_i -{\cal H}\big)\,,
}{eq:can11}
where the Hamiltonian density
\eq{
{\cal H} = \bar q_{i\,a}\, G_i^a
}{eq:can12}
is a sum over secondary constraints $G_i^a \approx 0$, as expected on general grounds.\footnote{The notation $\approx$ means `vanishing weakly' \cite{Dirac:1996}, i.e., vanishing on the surface of constraints.} They are given by
\begin{align}
G_1^a &=-\frac\ell 2 \,R^a +\frac \ell 2 \,\tilde R^a\,,\label{eq:G1}\\
G_2^a &=\frac\ell 2 \,\tilde D\la^a + 2\ell\, \tilde R^a \,,\label{eq:G2}\\
G_3^a &= -\frac\ell 2 \,D\la^a\,.\label{eq:G3}
\end{align}
We have introduced the following abbreviations
\eq{
R^a:=\big(\partial_1 A_2^a-\partial_2 A_1^a\big) + \frac12 \eps^a{}_{bc} \big( A_1^b A_2^c - A_2^b A_1^c\big)
}{eq:defR}
and
\eq{
D\la^a:= \big(\partial_1\la_2^a-\partial_2 \la_1^a\big) + \eps^a{}_{bc}\big(A_1^b\la_2^c-A_2^b\la_1^c\big)
}{eq:defD}
and similarly for $\tilde R$ and $\tilde D\la$, with $A$ replaced by $\tilde A$ in the definitions \eqref{eq:defR} and \eqref{eq:defD}, respectively.

We focus now on the first/second class properties of the constraints and on their Poisson bracket algebra. We have found 9 secondary constraints $G_i^a$. If all of them were first class then the physical phase space would be zero-dimensional, because each first class constraint eliminates two dimensions from the phase space, and the dimension of the phase space spanned by $q_i^a,p_i^a$ is 18.

\section{Constraint analysis}\label{se:4}

With the canonical Poisson bracket
\eq{
\{q_i^a(x),p_j^b(x^\prime)\} = \{q_i^a,p^{\prime\,b}_j\}=\de_{ij}\,\eta^{ab}\,\de^{(2)}(x-x^\prime)
}{eq:cb}
we can now calculate the Poisson brackets of the constraints $G_i^a$ with each other and with the Hamiltonian density. 
The latter,
\eq{
\{G_i^a,{\cal H}^{\prime}\} = \bar q_{j\,b}^\prime\, \{G_i^a,G_j^{\prime\,b}\}
}{eq:can14}
reduce to a sum over brackets between the secondary constraints. We calculate now these brackets explicitly.

To this end we express the secondary constraints \eqref{eq:G1}-\eqref{eq:G3} in terms of canonical coordinates and momenta:
\begin{align}
G_1^a &= -\partial_1 p_1^a -\partial_2 p_2^a - \eps^a{}_{bc} \,\big(\frac2\ell p_1^bp_2^c+\frac{1}{2\ell} p_2^bp_3^c+q_3^bp_1^c\big)\\
\widehat{G}_2^a &= G_2^a +G_3^a = -\partial_1 p_3^a-2\ell\,\partial_2q_3^a+\eps^a{}_{bc}\, p_i^b q_i^c \\
G_3^a &= -\frac\ell 2\big(\partial_1 q_2^a-\partial_2 q_1^a\big)+\eps^a{}_{bc}\,\big(p_1^bq_1^c+p_2^bq_2^c-\frac14 p_3^b q_1^c-\frac\ell 2 q_3^b q_2^c\big)
\end{align}
Note that instead of $G_2^a$ we use for convenience the linear combination $\widehat{G}_2^a = G_2^a +G_3^a$. Straightforward calculation obtains: 
\begin{align}
\{G_1^a,G_1^{\prime\,b}\} &= Z_{11}^{ab}\,\de^{(2)}(x-x^\prime) \label{eq:algebra1} \\
\{\widehat G_2^a,\widehat G_2^{\prime\,b}\} &= -\eps^{ab}{}_c \,\widehat G_2^c \,\de^{(2)}(x-x^\prime) \approx 0 \label{eq:algebra2} \\
\{G_3^a,G_3^{\prime\,b}\} &= -\eps^{ab}{}_c \, G_3^c \,\de^{(2)}(x-x^\prime)+Z^{ab}_{33}\, \de^{(2)}(x-x^\prime)  \label{eq:algebra3} \\
\{G_1^a,\widehat G_2^{\prime\,b}\} &= \eps^{ab}{}_c\,G_1^c\,\de^{(2)}(x-x^\prime) \approx 0 \label{eq:algebra4} \\
\{\widehat G_2^a,G_3^{\prime\,b}\} &= -\eps^{ab}{}_c\,G_3^c\,\de^{(2)}(x-x^\prime) \approx 0 \label{eq:algebra5} \\
\{G_1^a,G_3^{\prime\,b}\} &= -\eps^{ab}{}_c\,\big(G_1^c-\frac14\,\widehat G_2^c\big)\,\de^{(2)}(x-x^\prime) + Z_{13}^{ab}\,\de^{(2)}(x-x^\prime) \label{eq:algebra6}
\end{align}
We have used here the abbreviations
\begin{align}
Z^{ab}_{11} & = \frac{1}{2\ell}\,\big(p_2^ap_1^b-p_2^bp_1^a\big)\\
Z^{ab}_{33} & = \frac{\ell}{8}\,\big(q_2^a q_1^b-q_2^bq_1^a\big)  \\
Z^{ab}_{13} & = -\frac14\,\big(p_1^aq_1^b+p_2^aq_2^b\big) + \frac14 \,\eta^{ab}\,\big(p_1^cq_{1\,c}+p_2^cq_{2\,c}\big)
\end{align}
or, equivalently,
\begin{align}
Z^{ab}_{11} & = -\frac{1}{2\ell}\, \big(e^a \wedge e^b\big)_{12}\\
Z^{ab}_{33} & = -\frac{\ell}{8}\, \big(\la^a\wedge\la^b\big)_{12} \\
Z^{ab}_{13} & = \frac14\, \big(e^a\wedge\la^b\big)_{12}-\frac14\,\eta^{ab} \eta_{cd}\, \big(e^c\wedge\la^d\big)_{12}
\end{align}
If the quantities $Z^{ab}_{ij}$ were all vanishing then all secondary constraints would be first class. Since some of them are non-vanishing we have a certain number of second class constraints. Namely, not all entries of $Z_{11}^{ab}$ can vanish because this would lead to a singular Dreibein $e^a$. Similarly, not all entries $Z_{33}^{ab}$ can vanish because this would lead to a singular Lagrange multiplier 1-form $\la^a$. Since the algebra of constraints does not close we shall encounter ternary constraints from consistency requirements, namely the vanishing of the Poisson brackets \eqref{eq:can14}.

In the analysis below, the $9\times 9$-matrix 
\eq{
M_{ij}^{ab} :=  \int_{x^\prime} \!\!d^2x^\prime\,\{G_i^a,G_j^{\prime\,b}\} 
}{eq:can17}
evaluated on the surface of constraints will play a crucial role. First, note that before imposing 
the ternary constraints we can establish an upper bound on the dimension $2n$ of the
physical phase space in terms of the rank $r_M$ of $M_{ij}^{ab}$. We started with a phase space of dimension $18$ and accounted for $9$ constraints. The rank $r_M$ counts how many of these that are second class. Thus, before additional constraints are introduced we have
\eq{
2n \leq 18 - r_M - 2*(9-r_M) = r_M.
}{eq:can36}
Now we turn to the ternary constraints. We note that after imposing these we are done, since the 
consistency conditions analog to \eqref{eq:can14} arising from the $T_i^a$ do not generate quaternary 
constraints. Since the algebra \eqref{eq:algebra1}--\eqref{eq:algebra6} closes on $\delta$-functions, 
requiring vanishing of the brackets \eqref{eq:can14} is equivalent to requiring
\eq{
T_i^a := M_{ij}^{ab} \bar q_{j\,b} \approx 0.
}{eq:can35}
Because the ternary constraints $T_i^a$ contain the canonical partners of the primary constraints 
$\bar p_i^a$ complications arise, since some of the latter may lose their status as first 
class constraints. 
Thus we have to include the $\bar q_i^a$ as canonical variables, giving a phase space of dimension $36$ before imposing the constraints. We determine now the rank of the $27\times 27$ matrix
\eq{
\widehat M_{ij}^{ab} :=  \int_{x^\prime} \!\!d^2x^\prime\,\{C_i^a,C_j^{\prime\,b}\} 
}{eq:can20}
evaluated on the surface of constraints using the order $C_i^a=(\bar p_i^a, G_i^a, T_i^a)$. Because of 
\eqref{eq:can35} we have 
\eq{
\{T_i^a, \bar p_j^b \} = M_{ij}^{ab}, 
}{eq:can37}
and thus $\widehat M$ has the block form
\eq{
\widehat M \approx \left(\begin{array}{ccc}
{\mathbb O} & {\mathbb O} & -M^T\\
{\mathbb O} & M & B \\
M & -B^T & C \end{array} \right),
}{eq:can21}
where all the blocks are $9 \times 9$ matrices. The form of the non-vanishing matrices $B$ and $C$ is not needed for determining a lower bound for the rank of $\widehat M$. We can put all copies of $M$ and $M^T$ on lower triangular form by row-operations that do not spoil the block structure of \eqref{eq:can21}. This makes $\widehat M$ lower triangular with $3r_M$ non zero anti-diagonal elements. Thus, a lower bound for the rank $r_{\widehat M}$ of $\widehat M$ is $3r_M$. 

We are now in a position to count the number of linearly independent first- and second-class constraints. We have $r_{\widehat M}$ second class constraints. The total number of constraints is 9(primary) $+$ 9(secondary) $+$ 9(ternary) $=27$, but out of the nine ternary constraints $T_i^a$, only $r_M$ are linearly independent. 
This is so because of \eqref{eq:can35}.

Thus, the total number of linearly independent constraints is $9+9+r_M = 18 + r_M$, and $r_{\widehat M}$ of these are
second class. The dimension $2n$ of the physical phase space is therefore bounded by
\eq{
2n = 36 - r_{\widehat M} - 2*(18 + r_{M} - r_{\widehat M}) = r_{\widehat M} - 2r_M \geq r_M.
}{eq:result}
The two inequalities \eqref{eq:can36} and \eqref{eq:result} establish $2n = r_M$.

Thus, all that remains is to determine the rank of $M$. Using the order $G^a_i = (G_1^a, G_3^a, \hat{G}_2^a)$, $M$ has the block form
\eq{
M \approx \left(\begin{array}{cc}
A_{6\times 6} & {\mathbb O}_{6\times 3}  \\
{\mathbb O}_{3\times 6} & {\mathbb O}_{3\times 3} 
\end{array}\right)\,,\qquad A_{6\times 6}:=\left(\begin{array}{cc}
Z_{11} & Z_{13} \\
-Z_{13}^T & Z_{33}
\end{array}\right)\,.
}{eq:can16}
The block entries ${\mathbb O}_{x\times y}$ contain $x$ rows and $y$ columns of zeros. 
From \eqref{eq:can16} we deduce that the rank of the antisymmetric matrix $M^{ab}_{ij}$ must be either six, four or two. Its nine Eigenvalues $n_1\dots n_9$ are given by
\eq{
n_1\dots n_5 = 0\,,\qquad n_{6,7} = \pm \frac i4\, \big(e^a\wedge\la_a\big)_{12}\,,\qquad n_{8,9} = \pm \frac i4\, \sqrt{P} 
}{eq:can18}
Therefore its rank equals (at most) four, and not six as suggested by naive counting. The polynomial under the square root in the last expression in \eqref{eq:can18} is given by
\eq{
P=\frac{2}{\ell^2} \, \big(e^a\wedge e^b\big)_{12}\big(e_a\wedge e_b\big)_{12}+\frac{\ell^2}{8}\, \big(\la^a\wedge \la^b\big)_{12}\big(\la_a\wedge \la_b\big)_{12} + \big(e^a\wedge\la^b\big)_{12}\big(e_a\wedge\la_b\big)_{12}
}{eq:can19}
The rank of \eqref{eq:can16} is four in general and two if in addition the condition
\eq{
\big(e^a\wedge \la_a\big)_{12} = -p_1^a q_{1\,a} - p_2^a q_{2\,a} = 0 
}{eq:can33}
holds. Because of \eqref{eq:can83} on-shell we obtain
\eq{
e^a\wedge \la_a \propto e^a \wedge {\rm Ric}_a \propto R_{\mu\nu}\,dx^\mu\wedge dx^\nu = 0
}{eq:onshell}
where Ric$_a$ is the Ricci 1-form with respect to the Levi-Civita connection (we recall that on-shell torsion vanishes). Thus, the constraint \eqref{eq:can33} must hold on all classical solutions. Therefore, in the physically relevant sector, $2n = r_M = 2$.\footnote{It is possible, although not necessary, to impose \eqref{eq:can33} as a further constraint. This does not change anything essential about the counting procedure.} This completes our constraint analysis.\footnote{
As a consistency check we investigate now what happens when the torsion constraint is dropped in \eqref{eq:can1}. In the current formulation this can be achieved by imposing the constraints
\eq{
G_4^a=q_1^a\approx 0\,,\qquad G_5^a=q_2^a\approx 0\,,\qquad G_6^a=\bar q_1^a\approx 0\,.
}{eq:cancon1}
These constraints render the constraints $G_3^a$ and $T_i^a$ superfluous. Thus, we have now 24 linearly independent constraints, $\bar p_i^a$, $G_1^a$, $\widehat G_2^a$, $G_4^a$, $G_5^a$, $G_6^a$. The rank of the $24\times 24$ matrix analog to \eqref{eq:can21} turns out to be equal to twelve. Therefore, we have now twelve first class and twelve second class constraints, which eliminates all dimensions from the phase space. Thus no physical bulk degrees of freedom remain. This is the anticipated result.}

To summarize, the dimension of the physical phase space is two and therefore CCTMG exhibits one physical bulk degree of freedom, which at the linearized level coincides with the topologically massive graviton.

\section{Conclusions}\label{se:5}

In this paper we have reformulated cosmological topologically massive gravity at the chiral point as a Chern--Simons action plus the difference between two Einstein--Hilbert actions, see \eqref{eq:can10}. We have performed a canonical analysis and recovered the anticipated\footnote{A recent canonical analysis in the first order formulation \cite{Park:2008yy} obtains a 2-dimensional physical phase space `for each internal index $a$', i.e., a 6-dimensional physical phase space. This result disagrees with ours and with previous literature, but it is then interpreted as a single graviton degree of freedom, concurrent with our result. Correspondence with the author revealed that he found additional constraints after posting his e-print and that currently he is reconsidering the constraint algebra. Another recent analysis \cite{Carlip:2008} agrees with our results.
} result of one physical bulk degree of freedom, which at the linearized level corresponds to the topologically massive graviton.

We have also encountered sectors of our first order theory that are not related to the second order formulation with regular field configurations, but that may be worthwhile studying in their own right. For instance, if one imposes by hand the constraints $\bar q_1^a=0=\bar q_3^a$ then no ternary constraints arise, but the Dreibein and Lagrange multiplier fail to be invertible.

Finally, we mention that the Poisson bracket algebra of the secondary constraints \eqref{eq:algebra1}-\eqref{eq:algebra6} closes with $\de$-functions rather than with first derivatives thereof because of our gauge theoretic reformulation of CCTMG. The same happens in $1+1$ dimensions, where this feature was exhibited and exploited by Wolfgang Kummer and his `Vienna School' \cite{Grumiller:2002nm,Kummer:2005tx}.

\section*{Acknowledgments}

We thank Steve Carlip, Stanley Deser, Mu-In Park and Andy Strominger for correspondence. NJ thanks the CTP at MIT for its kind hospitality during parts of this work.
This work is supported in part by funds provided by the U.S. Department of Energy (DoE) under the cooperative research agreement DEFG02-05ER41360. DG is supported by the project MC-OIF 021421 of the European Commission under the Sixth EU Framework Programme for Research and Technological Development (FP6). The research of NJ was supported in part by the STINT CTP-Uppsala exchange program.


\end{document}